\documentclass[aps,prl,twocolumn,superscriptaddress]{revtex4-1}

\usepackage{float}
\usepackage{xcolor}
\usepackage{graphicx}% Include figure files
\usepackage{dcolumn}% Align table columns on decimal point
\usepackage{bm}% bold math
\usepackage{hyperref}% add hypertext capabilities
\usepackage{color}
\usepackage{amsmath}
\usepackage{amsfonts}
\usepackage{amssymb}
\begin{document}

%\preprint{APS/123-QED}

\title{Stable two-dimensional dumbbell stanene: a quantum spin Hall insulator} %Title of paper

% Force line breaks with \\

%\author{xxx}
%\affiliation{yyy}

\author{Peizhe Tang}
\affiliation{State Key Laboratory of Low-Dimensional Quantum Physics, Department of Physics, Tsinghua University, Beijing 100084, China}
\affiliation{Nano-Bio Spectroscopy group, Dpto.~F\'isica de Materiales, Universidad del Pa\'is Vasco, Centro de F\'isica de Materiales CSIC-UPV/EHU-MPC and DIPC, Av.~Tolosa 72, E-20018 San Sebasti\'an, Spain}

\author{Pengcheng Chen}
\affiliation{State Key Laboratory of Low-Dimensional Quantum Physics, Department of Physics, Tsinghua University, Beijing 100084, China}

\author{Wendong Cao}
\affiliation{State Key Laboratory of Low-Dimensional Quantum Physics, Department of Physics, Tsinghua University, Beijing 100084, China}

\author{Huaqing Huang}
\affiliation{State Key Laboratory of Low-Dimensional Quantum Physics, Department of Physics, Tsinghua University, Beijing 100084, China}

\author{Seymur Cahangirov}
\affiliation{Nano-Bio Spectroscopy group, Dpto.~F\'isica de Materiales, Universidad del Pa\'is Vasco, Centro de F\'isica de Materiales CSIC-UPV/EHU-MPC and DIPC, Av.~Tolosa 72, E-20018 San Sebasti\'an, Spain}

\author{Lede Xian}
\affiliation{Nano-Bio Spectroscopy group, Dpto.~F\'isica de Materiales, Universidad del Pa\'is Vasco, Centro de F\'isica de Materiales CSIC-UPV/EHU-MPC and DIPC, Av.~Tolosa 72, E-20018 San Sebasti\'an, Spain}

\author{Yong Xu}
\affiliation{Department of Physics, McCullough Building, Stanford University, Stanford, California 94305-4045, USA}
\affiliation{Institute for Advanced Study, Tsinghua University, Beijing 100084, People's Republic of China}

\author{Shou-Cheng Zhang}
\affiliation{Department of Physics, McCullough Building, Stanford University, Stanford, California 94305-4045, USA}
\affiliation{Institute for Advanced Study, Tsinghua University, Beijing 100084, People's Republic of China}

\author{Wenhui Duan}
\email{Corresponding author: dwh@phys.tsinghua.edu.cn}
\affiliation{State Key Laboratory of Low-Dimensional Quantum Physics, Department of Physics, Tsinghua University, Beijing 100084, China}
\affiliation{Institute for Advanced Study, Tsinghua University, Beijing 100084, People's Republic of China}
\affiliation{Collaborative Innovation Center of Quantum Matter, Beijing 100084, China}

\author{Angel Rubio}
\email{Corresponding author: angel.rubio@ehu.es}
\affiliation{Nano-Bio Spectroscopy group, Dpto.~F\'isica de Materiales, Universidad del Pa\'is Vasco, Centro de F\'isica de Materiales CSIC-UPV/EHU-MPC and DIPC, Av.~Tolosa 72, E-20018 San Sebasti\'an, Spain}


\date{\today}% It is always \today, today,

\begin{abstract}
We predict from first-principles calculations a novel structure of stanene with dumbbell units (DB), and show that it is a two-dimensional topological insulator with inverted band gap which can be tuned by compressive strain. Furthermore, we propose that the boron nitride sheet and reconstructed ($2\times2$) InSb(111) surfaces are ideal substrates for the experimental realization of DB stanene, maintaining its non-trivial topology. Combined with standard semiconductor technologies, such as magnetic doping and electrical gating, the quantum anomalous Hall effect, Chern half metallicity and topological superconductivity can be realized in DB stanene on those substrates. These properties make the two-dimensional supported stanene a good platform for the study of new quantum spin Hall insulator as well as other exotic quantum states of matter.
\end{abstract}


\maketitle
The low-dimensional group \uppercase\expandafter{\romannumeral4} elements, such as fullerene \cite{curl1997}, carbon nanotube \cite{charlier2007}, graphene \cite{neto2009} and Si/Ge nanowire \cite{rurali2010}, have attracted lots of attentions in the recent decades due to their exotic electronic properties. Especially for the two-dimensional (2D) graphene, its big success inspires the ``Graphene Age" in the fields of physics, chemistry and material science. For direct integration on silicon based electronic device, the search of other 2D group \uppercase\expandafter{\romannumeral4} materials has triggered enormous interest. Since the first proposal of low-buckled (LB) silicene \cite{cahangirov2009}, a dumbbell configuration of silicene \cite{HDS-silicene,*Ongun2013,*Kaltsas2013} and a flat germanene structure \cite{Angel2014} have been synthesized. In contrast to graphene and LB silicene, the bond lengthes in LB 2D layer of tin (called stanene) are much larger, and the relatively weak $\pi$-$\pi$ bonding cannot stabilize the planar configuration, resulting in instability of free standing LB stanene (see the Supplemental Material \cite{sm} for details).

Following the initial discovery of the quantum spin Hall (QSH) effect in the 2D topological insulator (TI) HgTe quantum wells \cite{Bernevig2006,Konig2007}, much attention has been focused on the search of 2D TI with larger energy gaps. Recently 2D stanene has been proposed as a promising material to realize the QSH insulator, with gaps as large as $0.3$ eV, with appropriate chemical functionalization \cite{Yong2013}.
Owing to time-reversal (TR) symmetry, spin-filtered helical edge states propagate along edges with dissipationless spin and charge currents, leading to promising potential applications in spintronics and the fault-tolerant quantum computation \cite{Qi2011,*Hasan2010}.

In this Letter, we propose a new structure of stanene with dumbbell units (DB) (free standing or supported on an insulating substrate). Based on \emph{ab initio} structure optimization, phonon dispersion and band structure calculations, we predict that this structure can be stable and behaves as 2D TI. The SOC band inversion occurs at the $\Gamma$ point and the gap size can be tuned by applying compressive strain. We further show that in the realistic growth on the substrate, for example the boron nitride (BN) sheet or reconstructed InSb(111) surface on which \emph{$\alpha$}-tin(111) thin films have been fabricated \cite{Tin1996,*fantini2000}, the topologically non-trivial properties of DB stanene are retained. As an attractive platform for TI applications, some exotic phenomena, such as the quantum anomalous Hall effect (QAHE) \cite{Haldane1988,*liu2008,*yu2010,*chang2013,*WuCJ2014}, Chern half metallicity \cite{hu2014} and TR invariant topological superconductivity \cite{qi2009,*Wang2014,*ZhangFan2013}, can be realized in the supported DB stanene system.

\begin{figure}[tbp]
   \centerline{ \includegraphics[clip,width=1.0\linewidth]{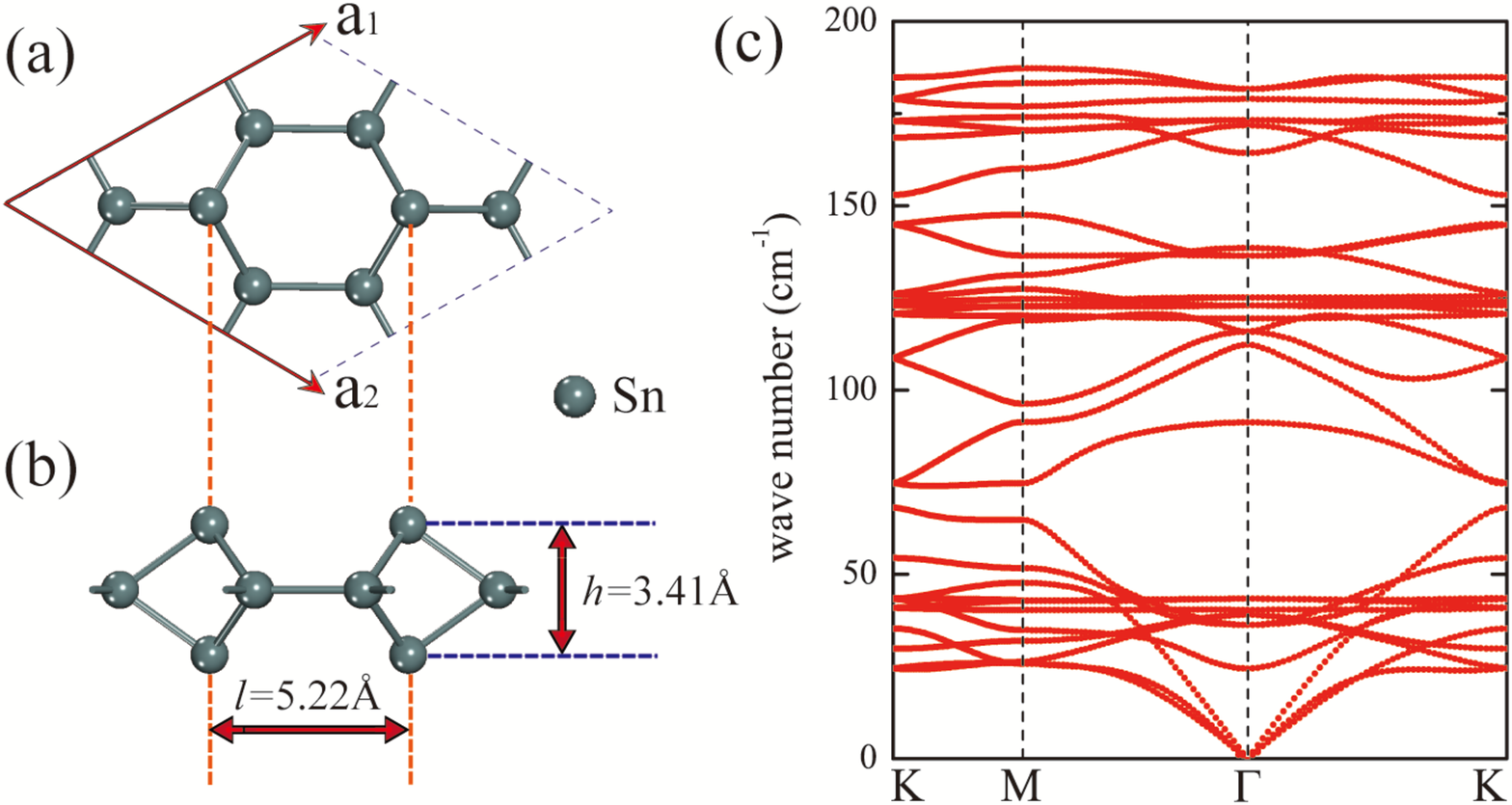}}
         \caption{(Color online) (a) Top view and (b) side view of the DB stanene structure. Black dashed lines mark the hexagonal lattice; $\bm{a_1}$ and $\bm{a_2}$ are the lattice unit vectors. $l$ is the distance between two DBs in a unitcell; and $h$ is the height of DB. (c) The phonon dispersion of the DB stanene structure.}
\label{fig:1}
\end{figure}

Density functional theory calculations were carried out by using the projector augmented wave method \cite{PAW1994,*PAW1999} and the generalized gradient approximation with Perdew-Burke-Ernzerhof type functional \cite{Perdew1996}, as implemented in the Vienna \textit{ab initio} simulation package \cite{Kresse1996}. Plane wave basis set with a kinetic energy cutoff of 300 eV was used. A slab model together with a vacuum layer larger than $20~\mathrm{\AA}$ was employed. During the structure optimization of DB stanene, all atomic positions and lattice parameters were fully relaxed, and the maximum force allowed on each atom was less than 0.01 eV/\AA. The Monkhorst-Pack \textit{k} points were 9$\times$9$\times$1. Phonon dispersions were obtained by using the frozen phonon method \cite{phonopy2008}.

Different from LB silicene, due to the larger bond length, the relatively weak $\pi$-$\pi$ bonding of LB stanene fails to stabilize the buckled configuration. Thus, we attend to add more Sn atoms forming DBs which can saturate the dangling bonds and stabilize the 2D structure. Then, a new stable configuration of stanene is discovered as shown in Figs. \ref{fig:1}(a) and (b). Meanwhile, as identified for silicene on Ag(111) surface, we succeed in achieving a similar DB structure of silicene which interacts with the substrate weakly \cite{HDS-silicene,*Ongun2013,*Kaltsas2013}. The optimized DB stanene has the hexagonal structure with space group D$_{6h}$ (P6/mmm) and ten Sn atoms in one unit cell. In this structure with lattice constant of $9.05~\mathrm{\AA}$, two DB units are formed at the para positions of a honeycomb ring and the other Sn atoms stay in the same plane. The height of DB [$h$ in Fig. \ref{fig:1}(b)] and the distance between the neighboring DBs [$l$ in Fig. \ref{fig:1}(b)] are $3.41~\mathrm{\AA}$ and $5.22~\mathrm{\AA}$, respectively. The calculated cohesive energy per Sn atom \cite{note1} for DB stanene is 0.18 eV larger than that of LB stanene, indicating that DB stanene is more stable. Furthermore, we calculate phonon dispersion for DB stanene [shown in Fig. \ref{fig:1}(c)]. It can be seen that the frequencies of all modes are positive over the whole Brillouin zone. This shows that DB stanene is thermodynamically stable and its stability does not depend on the substrate. So we expect that the free-standing DB stanene can be fabricated in the experiment.

\begin{figure}[tbp]
   \centerline{ \includegraphics[clip,width=1.0\linewidth]{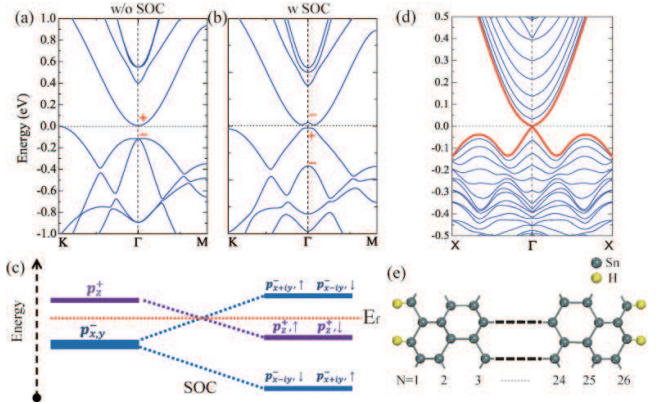}}
         \caption{(Color online) The band structures of DB stanene (a) without and (b) with SOC in the hexagonal lattice. Parities of the Bloch states at the $\Gamma$ point around the Fermi level are donated by $+$, $-$. (c) Schematic diagram of the band evolution of DB stanene driven by SOC for the orbitals around the Fermi level at the $\Gamma$ point. (d) The band structure and (e) schematic geometry of a $26$-DB stanene ZNR whose edges are passivated by hydrogen atoms. The Fermi level is set to zero.}
\label{fig:2}
\end{figure}

The band structures of DB stanene are shown in Figs. \ref{fig:2}(a) and (b). When the SOC is not included, it is a semiconductor with an indirect band gap. At the $\Gamma$ point, the top of valence band is mainly contributed by the antibonding $p_{x{\pm}iy}$ orbitals with fourfold degeneracy, and the bottom of conduction band is from the bonding $p_z$ orbital. When the SOC is included, the fourfold degenerate valence bands are split, and a band inversion occurs between $p_{x{\pm}iy}$ and $p_z$ orbitals with 40 meV non-trivial band gap around the $\Gamma$ point, as illustrated in Fig. \ref{fig:2}(c). The inverted states are labeled by $|p_{z}^{+}\rangle$ and $|p_{x{\pm}iy}^{-}\rangle$, where the superscript ($+$, $-$) denotes the parity. Because the inversion symmetry is present in DB stanene, the $\mathbb{Z}_{2}$ invariant can be calculated following the method proposed by Fu and Kane \cite{Fu2007}. We find the $\mathbb{Z}_{2}$ of DB stanene is $+1$, indicating that it is a QSH insulator. Due to the similar inversion mechanism with HgTe quantum wells \cite{Bernevig2006,Konig2007}, the low-energy physics of DB stanene around the $\Gamma$ point can be described by the Bernevig-Hughes-Zhang model \cite{Bernevig2006}. Interestingly, the gap at the $\Gamma$ point is found to be very sensitive to the in-plane strain, upon which the energy levels of inverted bonding and antibonding states move in opposite directions. Typically, for DB stanene under a $+ 3$\% compressive strain, the direct band gap and the gap at the $\Gamma$ point will increase to 64 meV and 342 meV, respectively. The evolution of gap size upon strain is consistent with hybrid functional calculations \cite{Heyd2003,*Heyd2006} (see Supplemental Material \cite{sm}).

The existence of helical edge states with the spin-momentum locking is one of the prominent features of 2D TI. In order to explicitly demonstrate the edge states for DB stanene, a zigzag nanoribbon (ZNR) with mirror symmetry is constructed with the edges passivated by hydrogen atoms [see Fig. \ref{fig:2}(e)]. Following the convention of graphene ZNR \cite{Son2006}, we classify the DB stanene-ZNR by the number of zigzag chains across the ribbon width. For avoiding interaction between the edge states, 26-DB stanene-ZNR is selected, of which the calculated band structure is presented in Fig. \ref{fig:2}(d). The helical edge states (marked by red lines) emerge from the bulk conduction band, cross at the $\Gamma$ point and enter the bulk valence band, exhibiting the topological nontrivial property. All the above results consistently indicate that DB stanene is an ideal 2D TI.

For the experimental realization, we note that the 2D group \uppercase\expandafter{\romannumeral4} element sheets can be achieved by various techniques, including mechanical exfoliation, chemical exfoliation, and molecular beam epitaxy (MBE). For example, high-quality graphene \cite{neto2009}, silicene \cite{vogt2012,*Arafune2013,*cahangirov2013} and Pb thin film \cite{Xue2004} have been successfully grown on different substrates based on MBE method. It is reasonable to expect that DB stanene can also be fabricated or transferred on the substrate using similar techniques. For the future applications, it is essential to find a proper substrate for DB stanene on which its exotic topological properties can be retained. Due to the close lattice structure, the hexagonal BN sheet and InSb(111) surface are good candidates. Especially on the reconstructed ($2\times2$) InSb(111) surface [hereafter referred to as InSb(111)-($2\times2$)], good-quality \emph{$\alpha$}-tin(111) thin films have been grown by using MBE method \cite{Tin1996,*fantini2000,Ohtsubo13,*Barfuss13}

\begin{figure}[tbp]
   \centerline{ \includegraphics[clip,width=1.0\linewidth]{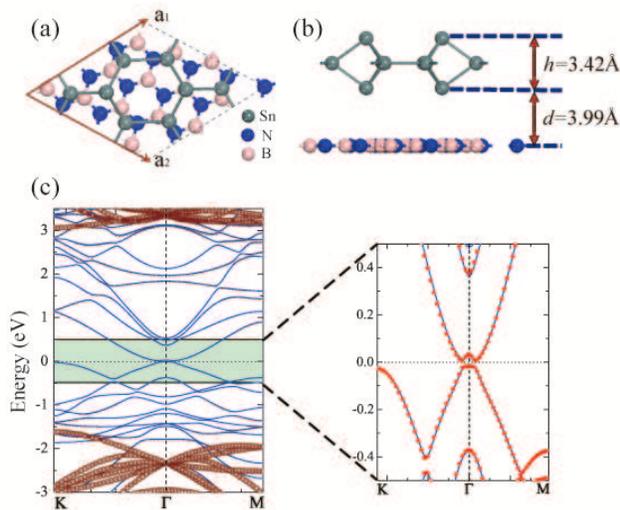}}
         \caption{(Color online) (a) Top view and (b) side view of DB stanene on the ($\sqrt{13}\times\sqrt{13}$) BN sheet. Black dashed lines mark the hexagonal lattice; $\bm{a_1}$ and $\bm{a_2}$ are the lattice unit vectors. $d$ is the distance between DB stanene and substrate. (c) The band structure of DB stanene on the ($\sqrt{13}\times\sqrt{13}$) BN sheet with SOC. The empty red dots stand for the contribution from the BN sheet. Inset shows the bands around the Fermi level, where red dots correspond to the bands of DB stanene without the substrate. The Fermi level is set to zero.}
\label{fig:3}
\end{figure}

As a 2D insulator with large band gap and high dielectric constant, the BN sheet has been used as substrate to grow graphene and assembled in many 2D stacked nanoelectronic devices \cite{britnell2012,*ju2014}. Here we use it as substrate to support DB stanene. Figs. \ref{fig:3}(a) and (b) show the geometrical structure of DB stanene on the ($\sqrt{13}\times\sqrt{13}$) BN sheet, where the lattice mismatch is only about $0.4\%$. After full relaxation, DB stanene almost retains the original structure with DB height of $h=3.42~\mathrm{\AA}$ [see Fig. \ref{fig:3}(b)]. And the distance between adjacent layers [$d$ in Fig. \ref{fig:3}(b)] is $3.99~\mathrm{\AA}$. The binding energy is about 0.04 eV per unit cell, showing the weak interaction between DB stanene and the BN sheet. The calculated band structure with SOC is shown in Fig. \ref{fig:3}(c): DB stanene on the BN sheet remains semiconducting; there is essentially no charge transfer between adjacent layers; and the states around the Fermi level are dominantly contributed by DB stanene. If we compare the bands of DB stanene with and without the BN substrate, which are marked by blue lines and red dots in the inset of Fig. \ref{fig:3}(c) respectively, little difference is observed. Evidently, DB stanene on the ($\sqrt{13}\times\sqrt{13}$) BN substrate is also a QSH insulator whose band inversion is not affected by the substrate.

\begin{figure*}[ht]
   \centerline{ \includegraphics[clip,width=0.9\linewidth]{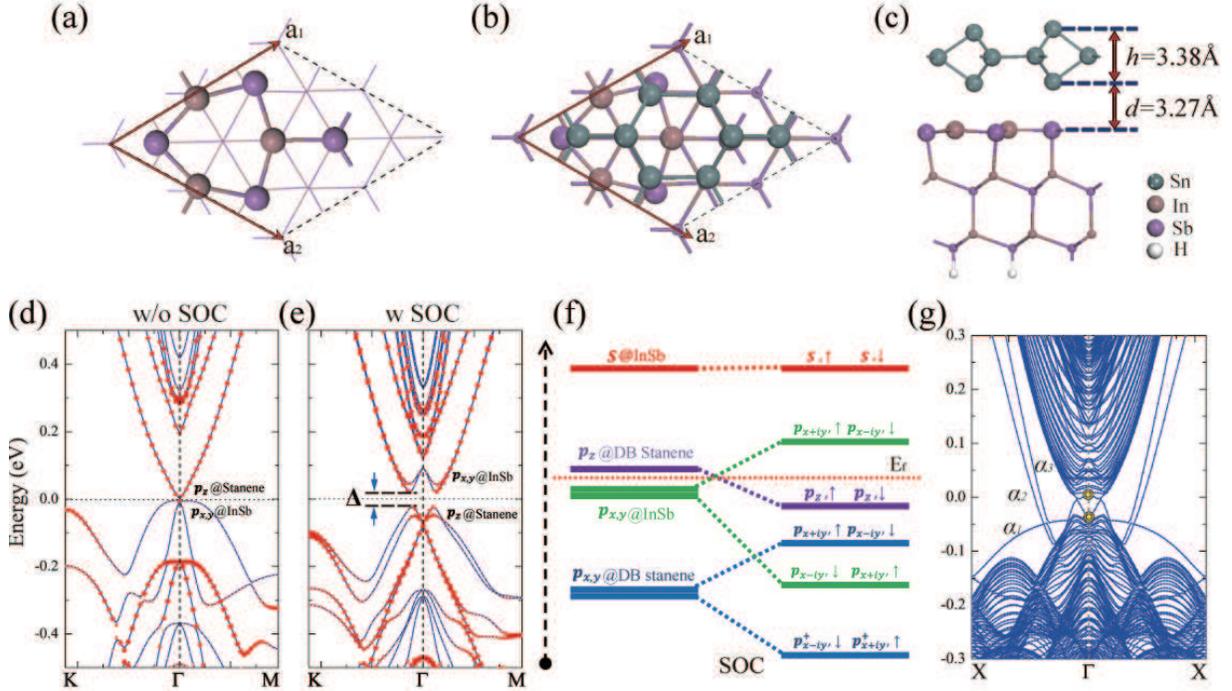}}
         \caption{(Color online) (a) Top view of the InSb(111)-($2\times2$) surface. (b) Top view and (c) side view of DB stanene on the InSb(111)-($2\times2$) thin film. Black dashed lines, $\bm{a_1}$, $\bm{a_2}$, $h$ and $d$ follow the same convention described in Fig. 3. The band structures of DB stanene on the InSb(111)-($2\times2$) thin film (d) without and (e) with SOC. The red dots stand for the contribution from DB stanene. (f) Schematic diagram of the band evolution at the $\Gamma$ point driven by SOC for the orbitals around the Fermi level. (g) The edge state of the DB stanene ribbon on the InSb(111)-($2\times2$) from \textit{ab initio} tight binding calculation. Yellow dots mark the Dirac points of edge states localized at different terminals. The Fermi level is set to zero.}
\label{fig:4}
\end{figure*}

In experiment, the InSb(111)-($2\times2$) surface with In-vacancy buckling has been used as an ideal substrate for the growth of \emph{$\alpha$}-tin(111) thin films \cite{Tin1996,*fantini2000}. Its geometrical structure is shown in Fig. \ref{fig:4}(a): one-quarter of In surface atoms are missing and the top In-layer is relaxed into an almost planar configuration. Due to the small lattice mismatch (about $1.2\%$), we use the InSb(111)-($2\times2$) slab as substrate to support DB stanene (note: the bottom layer of the slab is saturated by hydrogen atoms). During the structural optimization, the DB stanene and topmost four atomic layers of the InSb(111)-($2\times2$) thin film are allowed to fully relax and the rest atoms are frozen. The optimized structure is shown in Figs. \ref{fig:4}(b) and (c). Similar to the case on the BN substrate, DB stanene on the InSb(111)-($2\times2$) thin film retains the DB configuration with the height of $h=3.38~\mathrm{\AA}$. However, the distance between DB stanene and the substrate ($d=3.27~\mathrm{\AA}$) is smaller than that on the BN sheet and its binding energy (0.42 eV per unit cell) is one order of magnitude larger. These results suggest that the coupling between DB stanene and InSb(111)-($2\times2$) thin film is much stronger than in the case of the BN sheet.

The calculated band structures confirm the results discussed above, as presented in Figs. \ref{fig:4}(d) and (e). When the SOC is not included, the bottom of conduction band is mainly contributed by the $p_{z}$ orbital in DB stanene, which touches the fourfold degenerate states of $p_{x{\pm}iy}$ orbitals of the InSb(111)-($2\times2$) thin film at the $\Gamma$ point. And the interaction with the substrate pushes $p_{x{\pm}iy}$ orbitals of DB stanene downward to lower energy. When the SOC is included, the top of valence bands mainly contributed by the InSb(111)-($2\times2$) substrate is split and a band inversion occurs at the $\Gamma$ point with a large gap value of 161 meV. Due to the inversion symmetry breaking, the Rashba-like splitting is observable for both the conduction and valence bands, and the direct band gap marked by $\Delta$ is 40 meV [see Fig. \ref{fig:4}(e)]. In contrast to DB stanene on the BN sheet, the inverted states are $p_{z}$ orbital of DB stanene and $p_{x{\pm}iy}$ orbitals of the InSb(111)-($2\times2$) thin film, which are contributed by different parts of the heterostructure. Such an inversion mechanism [see Fig. \ref{fig:4}(f)] is similar to that of InAs/GaSb Type-\uppercase\expandafter{\romannumeral2} quantum well QSH insulator \cite{LiuCX2008,*Knez2011}.

Due to the inversion symmetry breaking in DB stanene supported on the InSb(111)-($2\times2$) thin film, the method proposed by Fu and Kane \cite{Fu2007} can not be used to calculate the $\mathbb{Z}_{2}$ invariant and a new method independent of the presence of inversion symmetry is needed. Following the work of Soluyanov and Vanderbilt \cite{Soluyanov2011-1,*Soluyanov2011-2}, we construct the maximally localized Wannier functions \cite{Marzari1997} from our \textit{ab initio} results and calculate the $\mathbb{Z}_{2}$ invariant from the evolution of the Wannier charge center (WCC) during the ``TR pumping'' process. An odd number of crossings for smooth WCC bands and the path of the largest gap center between WCCs is observed (see Supplemental Material \cite{sm}), thus the $\mathbb{Z}_{2}$ is $+1$ for DB stanene on the InSb(111)-($2\times2$) thin film, suggesting that the topological non-trivial property retains in the hybridized structure. To elucidate this finding, a ribbon model containing $30$ unit-cell-width is constructed, and the band structure from \textit{ab initio} tight binding calculation is shown in Fig. \ref{fig:4}(g). For the ribbon of DB stanene on the InSb(111)-($2\times2$) thin film, helical edge states contributed by different edges coexist with trivial edge states [$\alpha_{1,2,3}$ in Fig. \ref{fig:4}(g)]; and due to the breaking of mirror symmetry in the ribbon, the helical edge states are split in Fig. \ref{fig:4}(g) (see Supplemental Material \cite{sm}).

To conclude, based on first-principles calculations, we propose a new stable 2D structure for stanene with non-trivial topology. Its inverted band gap at the $\Gamma$ point is sensitive to compressive strain and thus a large value is possible to be achieved experimentally. Furthermore, we reveal the topological properties of DB stanene on different substrates for practical applications. DB stanene on the BN sheet retains the electronic properties of free-standing DB stanene due to the weak interaction with the BN sheet. DB stanene on the InSb(111)-($2\times2$) thin film is also identified as 2D TI without inversion symmetry, but its band inversion is also dependent on the substrate, which provides us an opportunity to manipulate the electronic properties of whole hetreostructure by using well-established technologies on \uppercase\expandafter{\romannumeral3}-\uppercase\expandafter{\romannumeral5} semiconductor. We expect to observe TR invariant topological superconductivity \cite{qi2009,*Wang2014,*ZhangFan2013} in the electrically gated system, and the QAHE \cite{Haldane1988,*liu2008,*yu2010,*chang2013,*WuCJ2014} and Chern half metal \cite{hu2014} via magnetically doping the InSb substrate \cite{csontos2005}. All of these make supported DB stanene as an ideal platform to study novel quantum states of matter, showing great potential for the future applications.

We thank B. H. Yan for stimulating discussions. We acknowledge support from the Ministry of Science and Technology of China (Grant Nos. 2011CB606405 and 2011CB921901) and the National Natural Science Foundation of China (Grant No. 11334006). AR, SC, and LX acknowledge financial support from the European Research Council Grant DYNamo (ERC-2010-AdG No. 267374) Spanish Grants (FIS2010-21282-C02-01), Grupos Consolidados UPV/EHU del Gobierno Vasco (IT578-13) and EC project CRONOS (280879-2 CRONOS CP-FP7). YX, and SCZ acknowledge the Defense Advanced Research Projects Agency Microsystems Technology Office, MesoDynamic Architecture Program (MESO) through the contract number N66001-11-1-4105 and FAME, one of six centers of STARnet, a Semiconductor Research Corporation program sponsored by MARCO and DARPA. The calculations were done on the ``Explorer 100'' cluster system of Tsinghua University.

%\bibliography{ref}

\begin{thebibliography}{50}
\bibitem [{not()}]{note1}%
  \BibitemOpen
  \href@noop {} {\ }\bibinfo {note} {The cohesive energy per Sn atom is defined
  as $(nE^{atom}-E_{total})/n$, where $E_{total}$ is the calculated total
  energy of stanene, $E^{atom}$ is the energy of Sn atom and $n$ is the number
  of Sn atoms in one unit cell.}\BibitemShut {Stop}%
\end{thebibliography}

\end{document}